\documentclass[
reprint,
superscriptaddress,
longbibliography,
 amsmath,amssymb,
 aps,
]{revtex4-2}

\usepackage{graphicx}
\usepackage{dcolumn}
\usepackage{bm}
\usepackage{physics}
\usepackage{siunitx}
\usepackage{color}
\usepackage{amsmath}
\usepackage[colorlinks=true, allcolors=blue]{hyperref}

\makeatletter
\def\maketitle{
\@author@finish
\title@column\titleblock@produce
\suppressfloats[t]}
\makeatother

\begin{document}

\preprint{APS/123-QED}

\title{Exploring Topological Effects in Thin-Film X-Ray Cavities }

\author{Hanns Zimmermann}
\email{hanns.zimmermann@unibw.de}
\affiliation{Julius-Maximilians-Universit{\"a}t W{\"u}rzburg, Institute for Theoretical Physics and Astrophysics, Am Hubland, 97074 W{\"u}rzburg, Germany}
\affiliation{Forschungsinstitut Code, Universität der Bundeswehr, Carl-Wery-Str. 18, 81739 München, Germany}
\affiliation{W{\"u}rzburg-Dresden Cluster of Excellence ct.qmat}
\author{Jonathan Sturm}%
\affiliation{Julius-Maximilians-Universit{\"a}t W{\"u}rzburg, Institute for Theoretical Physics and Astrophysics, Am Hubland, 97074 W{\"u}rzburg, Germany}
\affiliation{W{\"u}rzburg-Dresden Cluster of Excellence ct.qmat}
\author{Ion Cosma Fulga}
\author{Jeroen van den Brink}
\affiliation{Leibniz Institute for Solid State and Materials Research Dresden, Helmholtzstr. 20, 01069 Dresden, Germany
}%
\affiliation{W{\"u}rzburg-Dresden Cluster of Excellence ct.qmat}
\author{Adriana Pálffy}
\email{adriana.palffy-buss@uni-wuerzburg.de}
\affiliation{Julius-Maximilians-Universit{\"a}t W{\"u}rzburg, Institute for Theoretical Physics and Astrophysics, Am Hubland, 97074 W{\"u}rzburg, Germany}
\affiliation{W{\"u}rzburg-Dresden Cluster of Excellence ct.qmat}

\date{\today}

\begin{abstract}
Quantum control of single x-ray photons can be achieved using thin-film nanostructure cavities with embedded layers of resonant nuclei. 
Here, we design and theoretically investigate tailored cavity structures that implement a non-Hermitian version of the Su-Schrieffer-Heeger one-dimensional topological model. By tuning the geometry of the structure, different topological phases can be realized. We show that the presence of topological edge states can be identified in the reflectivity spectra of the thin-film cavities. Our findings pave the way for exploiting topological phases in x-ray quantum control. 
\end{abstract}

\maketitle

\emph{Introduction} ---
Hard x rays are widely used in many fields of natural sciences and industry exploiting their robustness, exceptional penetration depth, spatial resolution and detection efficiency \cite{xrayQO}. Given their short wavelength, x rays are less plagued by the diffraction limit and  can be focused down to the size of atoms, with great potential for ultimate miniaturization of photonic circuits \cite{Politi2008,Liao2012}, or  sensing with unprecedented spatial resolution \cite{Liao2014}. X-ray control is still far from the success achieved for optical and infrared photons \cite{Zeilinger2012}, but progress has been made in the past two decades in two directions: the commissioning of x-ray free electron lasers \cite{emma2010}, and the development of x-ray quantum optics platforms based on resonant interactions with M\"ossbauer nuclei, in which nuclear excitation and decay occur recoillessly. Thin-film nanostructures with embedded layers of resonant M\"ossbauer nuclei  are a remarkably clear quantum optics platform,  and have enabled a number of experimental achievements  such as observing the collective Lamb shift \cite{lambshift} or spontaneously generated coherences \cite{heeg2013vacuum},  realizing electromagnetically-induced transparency \cite{emtransparency}, demonstrating subluminal x-ray propagation \cite{slowlight2015},    or implementing collective strong coupling of single photons \cite{strongcoupling,Rabiosc}.

In the optical regime, special light propagation effects have been demonstrated in conjuncture with topology concepts. Since the discovery of the quantum Hall effect \cite{QuantumHall80,QuantumHall86}, topology has been applied to many different platforms ranging from classical \cite{topoacoust,topomech,topometa,toporeso} to quantum systems \cite{volovik2009,Eckardt2017,Cooper2008,Dalibard2011}. In photonics,  the use of carefully designed wavevector-space topologies allows the creation of interfaces that support new states of light \cite{Lu2014,TopoPhotonics}. Systems such as photonic crystals \cite{Haldane2008,Wang2008,Xiao2014} and coupled waveguides \cite{Rechtsman2013a,Rechtsman2013b,Ghareh2018} were used to implement  topological models like the one-dimensional Su-Schrieffer-Heeger (SSH) chain \cite{SSHPaper} or 
two-dimensional quantum Hall systems \cite{QuantumHall80}, leading to new applications, for instance several implementations of topological lasers \cite{Parto2018,stjean2017,Zhao2018,vcsels2021}. Quantum optics analogs of interacting topological systems were explored in  atomic emitter arrays of different geometries  \cite{Perczel2017, Perczel2020}.

In this Letter, we investigate theoretically how topological concepts can be transferred to the realm of x-ray photons. We show that specially designed nanostructures with multiple layers of M\"ossbauer nuclei can implement a non-Hermitian version of the well-known  SSH  model \cite{SSHPaper}. The couplings between different layers of M\"ossbauer nuclei,  generally  long-ranged, can be engineered by inserting in between additional layers of attenuating materials with high atomic number $Z$ in alternating thicknesses. Our results demonstrate the existence of topological edge states whose presence can be experimentally identified in the x-ray reflectivity spectra. The calculated corresponding topological invariant of the system, the winding number, confirms the emergence of a topological phase transition.  Our findings open up a new area of research in x-ray quantum optics  exploiting the robustness and protection specific for topological effects, towards x-ray propagation control. 

\emph{Stacked Thin-Film Cavities} ---
Thin-film x-ray cavities are multi-layer structures with alternating nm-scale-thick layers of varying atomic number $Z$.  In the simplest design, the center consists of the guiding layer made of material with low $Z$, i.e., low electron density, surrounded by high electronic density layers (corresponding to high $Z$) acting as mirrors. Inside the guiding layer, one can embed a layer of $^{57}\mathrm{Fe}$ nuclei, which have a M\"ossbauer magnetic-dipole transition between the ground state and the first excited state at $\SI{14.413}{keV}$. When illuminating in grazing incidence, i.e., at mrad incidence angles, incoming x rays can couple evanescently to the cavity. The  reflection of the incident light is substantially suppressed
at particular resonant incidence angles, as the x-ray  radiation  couples into a guided mode in the cavity. The latter  can drive  the nuclear transition in the  $^{57}\mathrm{Fe}$ layer, leading to repeated recoilless absorption and reemission of the resonant x-ray quanta  \cite{Ralf2004}. 

Our aim is to design and investigate a multi-layer nanostructure where the embedded $^{57}\mathrm{Fe}$ layers play the role of the SSH chain sites. The SSH model refers  to a one-dimensional lattice model with topological features, originally introduced to describe polyacetylene \cite{SSHPaper}. The underlying generic chain consists of 
identical unit cells, each with two sites $A$ and $B$, coupling only to their nearest neighbors, as illustrated in Fig.~\ref{cavity} (a). Depending on whether the intracell $(v)$ or the intercell $(w)$ hopping amplitude (both real-valued) is larger, we distinguish between a trivial or a topological phase, respectively \cite{Asboth2016}. In the topological phase, the system presents two so-called \emph{edge states}: zero-energy eigenstates that are concentrated entirely at the boundaries of the SSH chain. 
In order to implement a generalized SSH model in our system, we  require several layers of M\"ossbauer nuclei, with adjustable inter-layer couplings.
 To this end, we investigate a structure consisting of stacked  single $^{57}\mathrm{Fe}$-layer cavities with the same core structure, as exemplified in Fig.~\ref{cavity} (b). The attenuating properties of the separating high-$Z$ layers offer means of control over the inter-layer coupling without having to change the positions of the nuclear ensemble inside the guiding layers of each cavity. In the following, we proceed to  theoretically model this system, revealing its similiarities and differences to the traditional SSH chain. 
 
\begin{figure}
    \includegraphics[width=\linewidth]{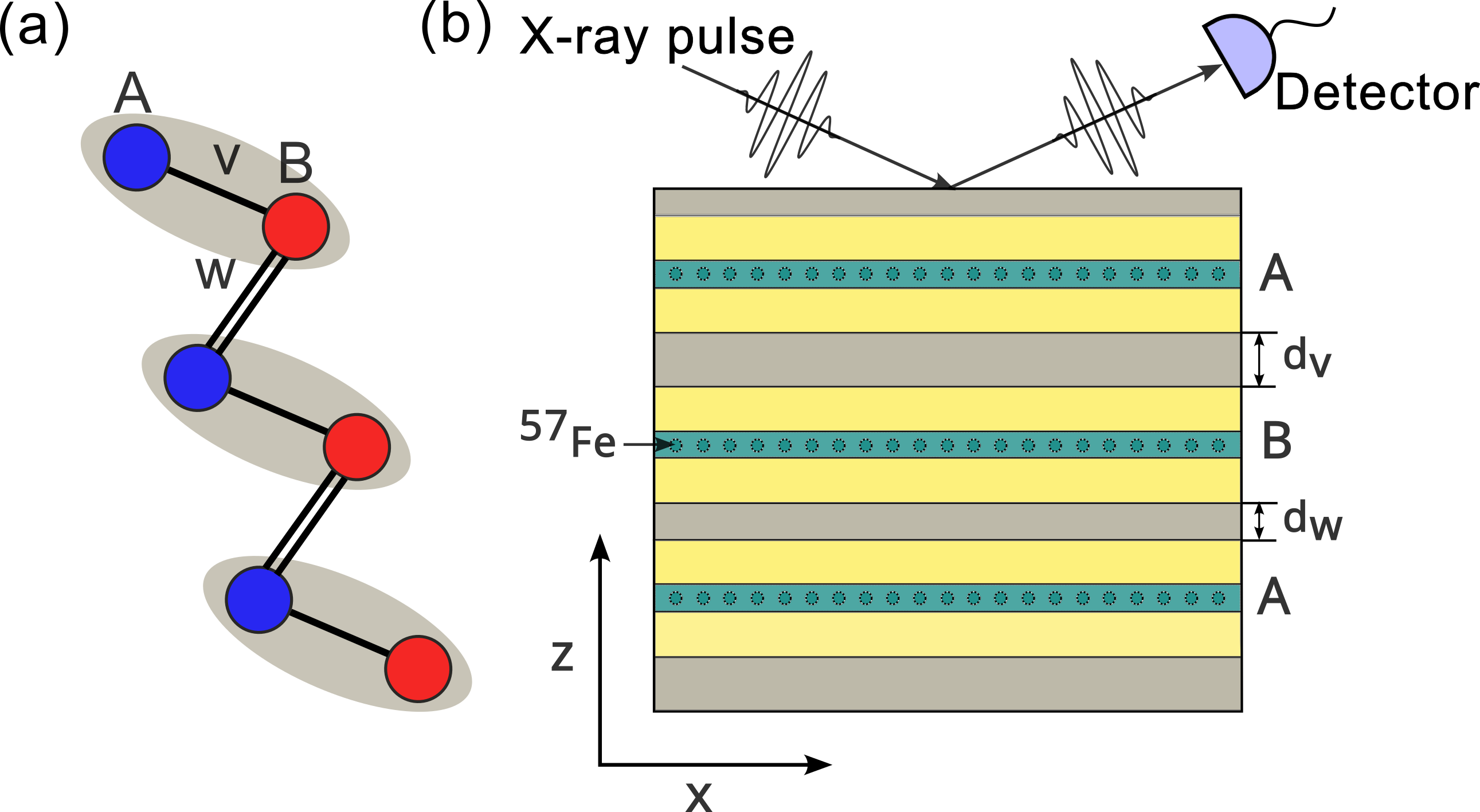}
    \caption{(a) Illustration of a SSH chain of unit cells (gray), each containing two sites $A$ (red)  and $B$ (blue), which are only coupled to their nearest neighbors with intracell (intercell) couplings $v\, (w)$. (b)
    Stack of three thin-film x-ray cavities, each containing one layer with $^{57}\mathrm{Fe}$ nuclei (green). The nuclear layers are surrounded by the guiding layer (yellow) made out of a low-$Z$ material like carbon. The entire structure is capped by layers of high-$Z$ material, for instance platinum (gray), used  also to separate the individual cavities with layers of alternating thicknesses $d_{v}$ and $d_{w}$.    
    \label{cavity}
    }
\end{figure}

We employ a quantum optical model based on the classical electromagnetic Green’s
function  \cite{standardxraynuc,Lentrodt2020}. For the scenario of  a single resonant x-ray photon in the incoming pulse, appropriate for synchrotron radiation, the effective non-Hermitian Hamiltonian for a system of $M$ nuclear layers reads (with $\hbar=$1) $\hat{H}_{\mathrm{eff}} = \hat{H}_{\mathrm D}+\hat{H}_{\mathrm{I}}$, with the coherent  driving term
\begin{equation}
\label{Hdrive}
    \hat{H}_{\mathrm{D}} =  -\sum_{j=1}^{M}\Bigl(\Omega_{j}\hat{S}_{j}^{\dagger}+\Omega_{j}^{\ast}\hat{S}_{j}\Bigr)\, ,
\end{equation}
and the nucleus-nucleus interaction Hamiltonian
\begin{equation}
       \label{Hnuc}
    \hat{H}_{\mathrm{I}} =- \Bigl(\Delta + \mathrm{i}\frac{\Gamma_{0}}{2}\Bigr)\sum_{j=1}^{M}\hat{S}_{j}^{\dagger}\hat{S}_{j}
     -\sum_{j,l=1}^{M}\Bigl(J_{jl}+\mathrm{i}\frac{\Gamma_{jl}}{2}\Bigr)\hat{S}_{j}^{\dagger}\hat{S}_{l}\, .
\end{equation}
In Eq.~\eqref{Hdrive}, $\Omega_j$ is the Rabi frequency of the $j$th $^{57}\mathrm{Fe}$ layer and $\hat{S}_{j}$ its collective nuclear spin-wave operator describing the collective excitation of the M\"ossbauer nuclei therein \cite{standardxraynuc,Lentrodt2020}. For the $j$th layer at position $z_j$ containing $N$ $^{57}\mathrm{Fe}$ nuclei, the Rabi frequency can be written as 
\begin{equation}
\label{Rabi}
        \Omega_j= \sqrt{N}\, \mathbf{m}^\ast\cdot \mathbf{B}_{\mathrm{1D}}(z_{j})\, ,
\end{equation}
where $\mathbf{m}$ is the magnetic dipole moment matrix element associated with the nuclear transition and $\mathbf{B}_{\mathrm{1D}}(z)$ is the magnetic component of the electromagnetic field scattered by the cavity, as described in the Supplementary Material (SM) \cite{supplmat}. In Eq.~\eqref{Hnuc},  $\Delta=\omega-\omega_0$ is the detuning between the incident x-ray field  and the nuclear transition characterized by angular frequencies $\omega$ and $\omega_0$, respectively, while $\Gamma_{0}$ denotes the spontaneous decay rate of the nuclear excited state of a single nucleus. The inter-layer couplings $J_{jl}$ and $\Gamma_{jl} $  can be related to the Green's function of the system by
\begin{equation}
\label{couplings }
        J_{jl}+\frac{\mathrm{i}}{2}\Gamma_{jl}=\frac{N\mu_{0}\omega^{2}}{ A} \mathbf{m}^{\ast}\cdot  \mathbf{G}(z_{j},z_{l},\omega,\mathbf{p}^{\mathbf{\rho}})\cdot \mathbf{m}\, ,
\end{equation}
where we have considered all layers to contain $N$ nuclei and have area $A$ in the $(x,y)$ plane. Furthermore, $\mu_0$ denotes the vacuum permeability. Exploiting the translational invariance of the system, the thin-film
x-ray cavity is treated as a quasi-1D structure along the $z$ direction, with $\mathbf{\rho}=(x,y)$ and $\mathbf{p}^{\mathbf{\rho}}$ being the transversal component of the incident x-ray wave vector $\mathbf{p}$. 
Then, the one-dimensional classical electromagnetic  Green's function $\mathbf{G}(z_{j},z_{l},\omega,\mathbf{p}^{\rho})$ for the $z$ direction can be obtained by solving numerically the classical light scattering problem in layered structures via the transfer matrix method \cite{GFTomasz,GFJohansson}. The Green's function is also employed to determine  $\mathbf{B}_{\mathrm{1D}}(z)$ (see SM \cite{supplmat}).

Typically, the Green's function for multi-layer cavities can have guided modes stretching in the $z$ direction across the entire structure. Such a field structure  leads to  all-to-all couplings between the $^{57}\mathrm{Fe}$ layers, with the inter-layer coupling depending in a non-trivial way on the positions of each respective pair of layers. 
In order to implement the SSH model, we are interested to suppress couplings beyond nearest-neighbor. This can be achieved by the attenuating high-$Z$ layers. Instead of stretching across the entire structure, the Green's function will be localized at its source cavity and only leak into its neighboring cavities. The coupling between two neighboring  $^{57}\mathrm{Fe}$ layers becomes a function of the width of the separating high-$Z$ layer $J_{j,j\pm 1}=J(d_{ v/w})$.

Using alternating widths of the attenuating high-$Z$ layer $d_{v}$ and $d_{w}$, we can mimic the inter- and intra-cell hopping amplitudes of the SSH model, with a number of differences. First, the real-valued spin-exchange inter-layer couplings $J_{jl}$ are accompanied by imaginary terms modeling the decay rates $\Gamma_{jl}$. Thus, our system  described by the nucleus-nucleus interaction Hamiltonian $\hat{H}_{\mathrm{I}}$ in Eq.~\eqref{Hnuc} approximates a non-Hermitian SSH chain which is coupled to the external cavity field described by the Rabi frequency terms. Second, our model  involves non-zero self-coupling terms  $J_{ll}$, which is not the case in the SSH model. Technically, the chiral symmetry should be broken, which generally would imply that there are no non-trivial topological phases in one dimension \cite{Ryu_2010}. However, provided that all self-coupling terms are equal, they contribute only via a term proportional to the identity operator to the Hamiltonian,  such that topological effects are not ruled out \cite{Wang2018}. Third, weak  beyond nearest-neighbor couplings between the iron layers may persist despite our design efforts.

We proceed to investigate the spectrum and  the eigenstates of the interaction Hamiltonian in Eq.~\eqref{Hnuc}, for the case of zero detuning $\Delta=0$. Numerically, we consider a multi-layer structure comprised of ten stacked single-layer cavities, separated by platinum layers of alternating widths   $d_{v}$ and $d_{w}$ as illustrated for only three cavities in Fig.~\ref{cavity} (b). The cores of all single-layer cavities have the same structure  ($\SI{19.5}{nm}$ C)/($\SI{1}{nm}$ $^{57}\mathrm{Fe}$)/($\SI{19.5}{nm}$ C), while the entire stack is capped by $\SI{2.5}{nm}$ thick platinum layers on both  top and  bottom. To keep the multi-layer structure perfectly symmetric, we assume it surrounded by air, with  neither super- nor substrate. Asymmetries in the latter do not affect the emergence of edge states, but lift their degeneracy. The incoming x-ray pulse illuminates the structure at a fixed incidence angle $\varphi=\SI{2.4067}{mrad}$, corresponding to the vicinity of the first cavity mode for a wider range of values  $d_{v}$ and $d_{w}$. To investigate a wide range of possible couplings, we keep $d_{w}$ fixed at $\SI{3.5}{nm}$ and vary the width $d_{v}$ in the interval  $0.5\le d_{v}/d_{w}\le 2$.

The real part of the numerically calculated eigenstates is presented in Fig.~\ref{edge-states}~(a). While for $d_{v}/d_{w}<1$ two separated bands are visible, in the vicinity of $d_{v}=d_{w}$ two states, each from one of the bands, approach and merge into a common mid-gap state. Due to the non-zero self coupling, these mid-gap states do not have zero energy as typical for the standard SSH model. The mid-gap state energy is given by the (equal) self-couplings of the outmost iron layers, and could be tuned  and even made to vanish by choosing an appropriate detuning $\Delta$. Note that, in contrast to the SSH model, the eigenvalues have also an imaginary part, not plotted here. The spatial localization of the corresponding eigenstates is shown in Fig.~\ref{edge-states}~(b) for two ratios $d_{v}/d_{w}$. We notice that while for a subunitary ratio $d_{v}/d_{w}=0.8$, the ten eigenstates are more or less equally spread over the ten $^{57}\mathrm{Fe}$ layers, two edge states localized at the outmost iron layers [corresponding to the mid-gap states in Fig.~\ref{edge-states}~(a)] appear in the case of 
$d_{v}/d_{w}=1.4$, as we would expect from the SSH model.  

\begin{figure*}

        \includegraphics[width=0.31\linewidth]{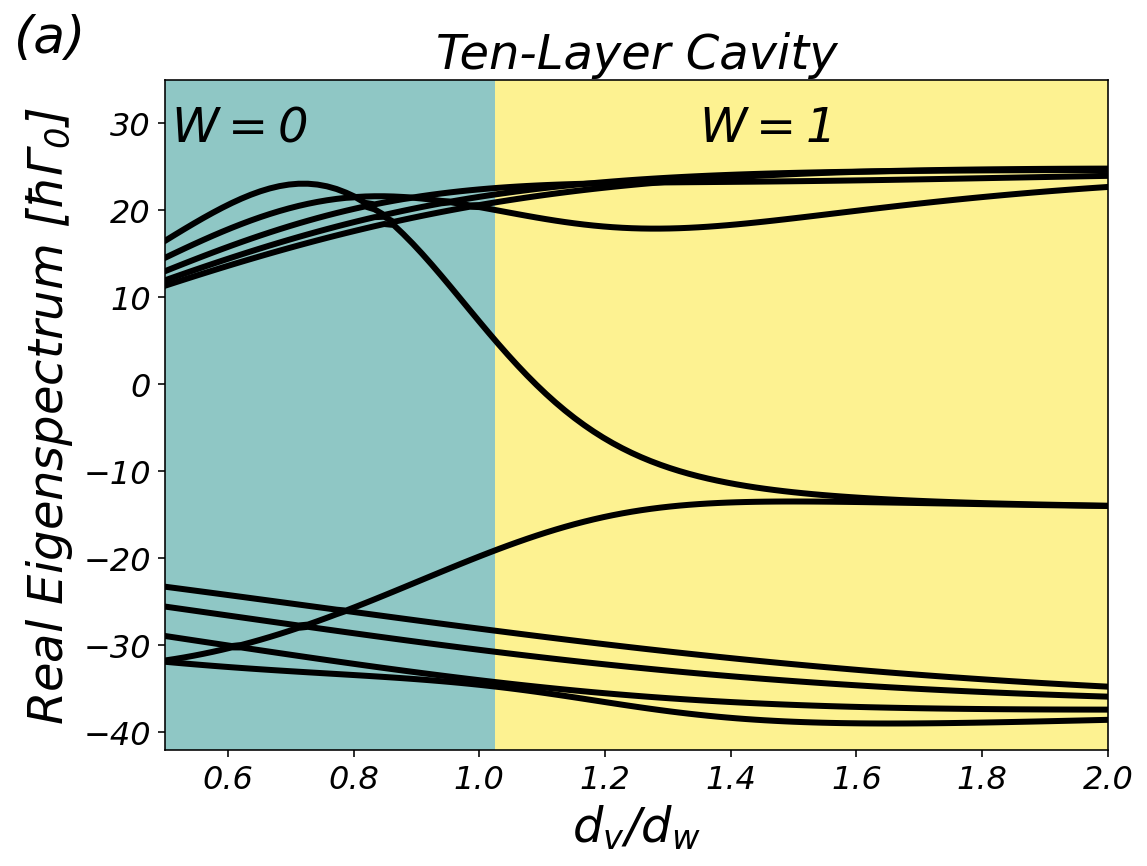}   
        \includegraphics[width=0.31\linewidth]{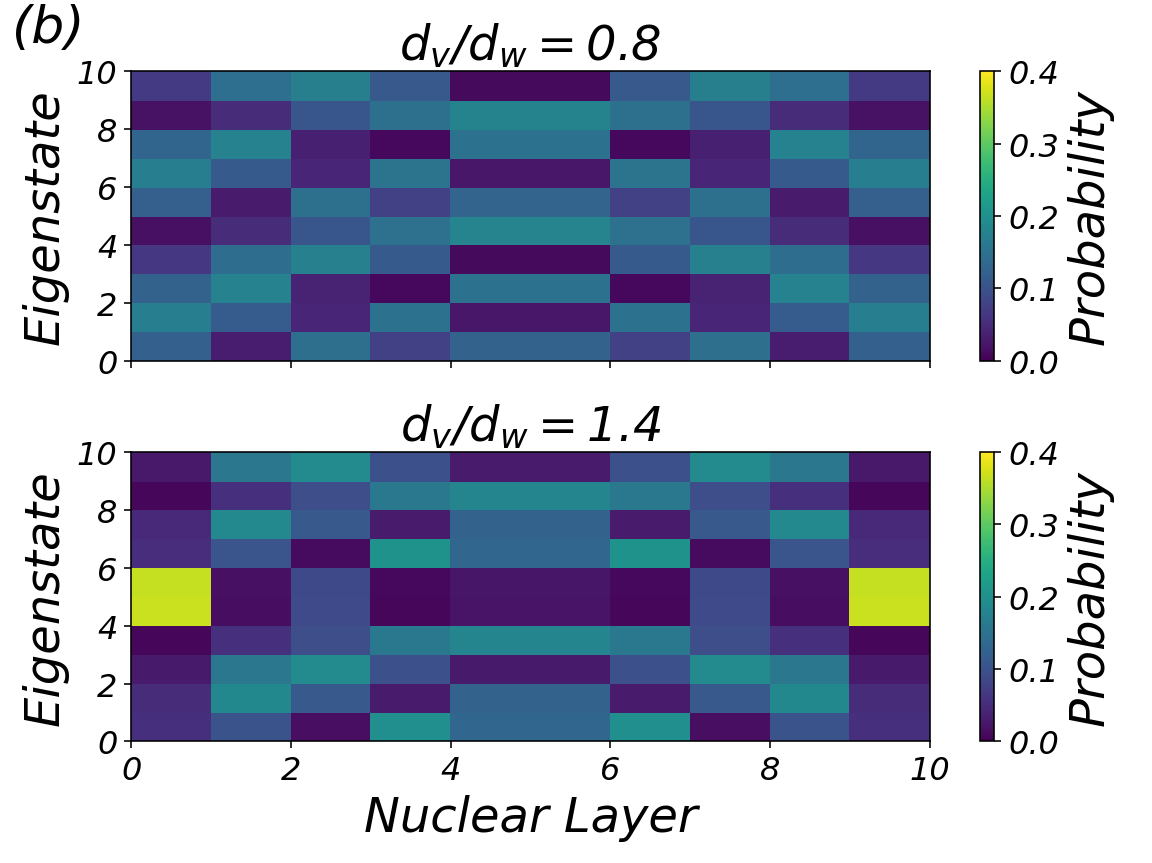}
         \includegraphics[width=0.31\linewidth]{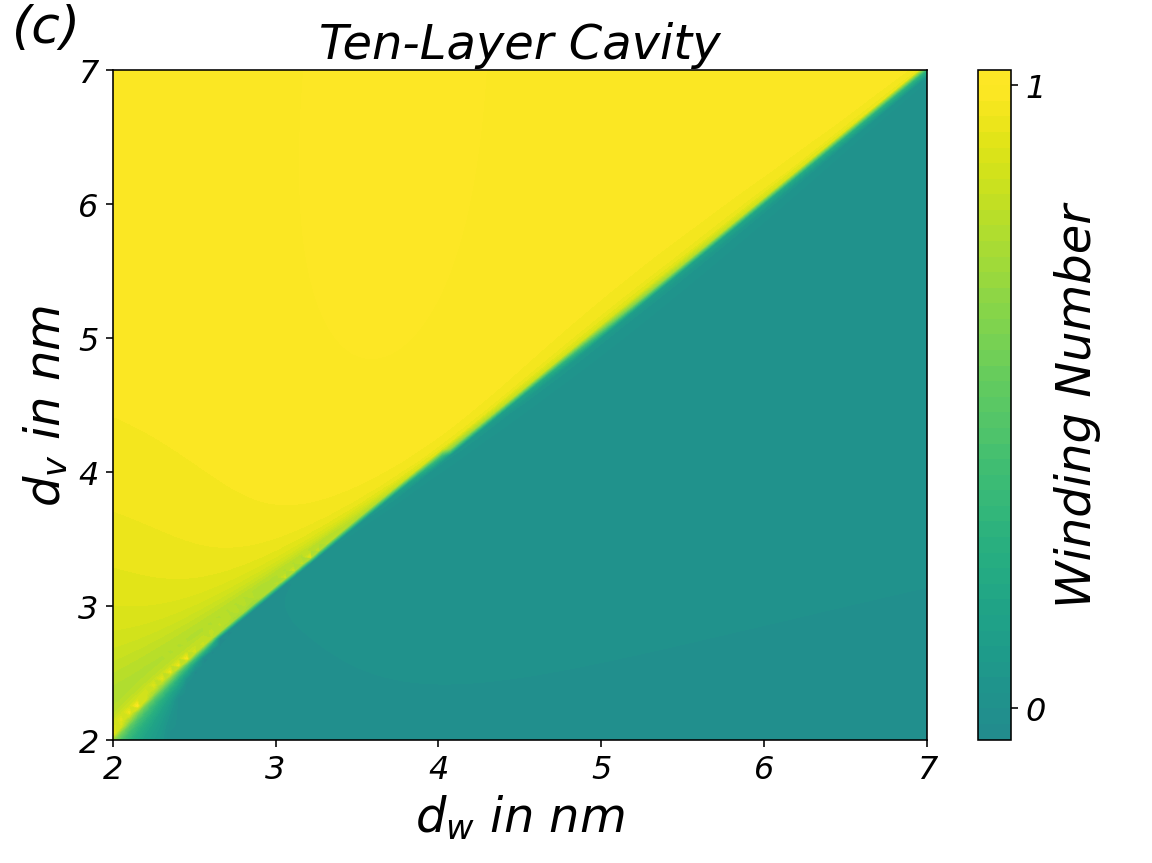}
    
    \caption{ \label{edge-states}
    (a) Real part of the eigenenergies of $\hat{H}_{\mathrm{I}}$ for $\Delta=0$ as a function of the width $d_{v}$ for a system of ten stacked single-layer cavities with a fixed incidence angle $\varphi=\SI{2.4067}{\mathrm{mrad}}$ and fixed width $d_{w}=\SI{3.5}{nm}$. The full Hamiltonian is given in the SM \cite{supplmat}.
    The shaded areas depict the corresponding calculated winding numbers.
    (b) Nuclear layer localization of the corresponding eigenstates of $\hat{H}_{\mathrm{I}}$ 
    for the topologically trivial case with ratio $d_{v}/d_{w}=0.8$ (upper) and the topological case with a ratio $d_{v}/d_{w}=1.4$ (lower). 
    (c)  Corresponding winding number as a function of the thicknesses $d_{v}$ and $d_{w}$ of the attenuating layers for a fixed x-ray incidence angle at $\varphi=\SI{2.4067}{mrad}$.  }
\end{figure*}


\emph{Topological Invariant} ---
 According to the bulk-boundary correspondence \cite{Asboth2016}, the existence of edge states is connected to the winding number $W$, which is a topological invariant, i.e., it remains unchanged while the system undergoes adiabatic deformations. For the standard SSH model, the winding number can take only two different integer values, $W=0$ for  the trivial and $W=1$ for the topological case, respectively. The winding number can be calculated using Bloch's theorem to deduce a bulk Hamiltonian $\hat{\mathcal{H}}(k)$ under the assumption of periodic boundary conditions, where $k$ denotes the crystal quasi-momentum. For a non-Hermitian SSH model, the 
winding number  is defined as \cite{nonhermssh}
\begin{equation}
    W=\frac{\mathrm{i}}{\pi}\int_{\mathrm{BZ}}\bra{\lambda_{k}^{n}}\frac{\partial}{\partial k}\ket{\psi_{k}^{n}} dk, 
\end{equation}
where the integral is taken over the 1D Brillouin zone BZ, $n$ denotes the band index and $\ket{\lambda}$ ($\ket{\psi}$) are the left (right) eigenstates of the Bloch Hamiltonian $\hat{\mathcal{H}}(k)$.

 To calculate the winding number for our multi-layer system,  we derive the Bloch Hamiltonian corresponding to the  nuclear-interaction Hamiltonian $\hat{H}_{\mathrm{I}}$ (for $\Delta=0$)  under the assumption of periodic boundary conditions. We enforce the latter by discarding the four outmost iron layers and considering only the bulk layers (see matrix example in SM \cite{supplmat}). The latter have approximately the same self-coupling value, such that the main diagonal in the Bloch Hamiltonian has no impact in the value of the winding number \cite{Wang2018,Pocock2018}. The resulting winding number for our idealized cavity stack is shown as shading for the corresponding ratio  $d_{v}/d_{w}$ in Fig.~\ref{edge-states}~(a) and more generally  
 for varying configurations of the thicknesses $d_{v}$ and $d_{w}$ in Fig.~\ref{edge-states}~(c). We consider only thicknesses larger than 2 nm since otherwise long-range couplings become too strong to justify our model.
 The case presented in Fig.~\ref{edge-states}~(a) corresponds to a vertical cut in Fig.~\ref{edge-states}~(c) at $d_{w}=3.5$~nm.
 As a general trend, for $d_{v}>d_{w}$, i.e., intracell coupling weaker than the intercell coupling, the system is in a topological phase with the winding number $W=1$. The  system remains in a trivial phase with winding number $W=0$ for the opposite case $d_{v}<d_{w}$. For the case of $d_{w}=3.5$~nm for which the real part of the eigenenergies is presented in Fig.~\ref{edge-states}~(a), the phase transition occurs at $d_{v}/d_{w}=1.02$.  

Figure \ref{edge-states} (c) shows that the transition between the two phases predicted by the winding number occurs approximately at $d_{v}/d_{w}=1.0$ as expected in the SSH model. Around the phase transition, the Block bands become degenerate, leading to an ill-defined winding number with non-integer values. Also, thinner Pt layers (small 
 $d_{v}$ and $d_{w}$)  lead to non-negligible long-range couplings and therefore next-to-nearest neighbor couplings, which break translational symmetry.
While our results in Figs.~\ref{edge-states} generally confirm the bulk-boundary correspondence for our system, the change in winding number does not coincide exactly with either the emergence of the edge states or the merging of their respective energies. The energies of the two edge states change gradually and  merge at a 
larger   $d_{v}/d_{w}$ ratio than the one where the winding number changes. This is the consequence of the interplay of two effects.  First, depending on the values of  $d_{v}$ and $d_{w}$, non-negligible long-range couplings beyond nearest-neighbor may be present. This is true even for thicker attenuating platinum layers. Second, the considered chain is very short. Our numerical results show that for a longer chain, the two edge states merge very close to the winding number jump.


\emph{Experimental Signature} ---
Let us now focus on how to identify the topological phase of the multilayer cavity system in an experiment. The observable of choice is the cavity reflectivity, i.e., the  squared ratio of the outgoing field caused by electronic and nuclear scattering and the incident x-ray field, usually measured in frequency or time domain. We use the so-called input-out formalism 
 \cite{standardxraynuc,inoutQO} and the already introduced Green's function 
 to express the outgoing magnetic field in terms of the incoming field.
Thereby, we use biorthogonal quantum mechanics \cite{Brody_2014} to express the outgoing field in terms of the eigenstates of the Hamiltonian $\hat{H}_{\rm  I}$ from Eq.~(\ref{Hnuc}) for zero detuning. 
For a system of $M$ identical layers with embedded $^{57}\mathrm{Fe}$ nuclei and assuming the incoming field at grazing incidence with linear polarization such that we can neglect the vector character of the magnetic field, the outgoing field at the position of the source in the top layer $z_{\mathrm{src}}$ reads (see SM \cite{supplmat} for derivation)
\begin{equation}
\label{inout}
\begin{split}
    B_{\mathrm{out}}(\Delta)=&B_{\mathrm{cav}}(z_{\mathrm{src}})\\
    &-\mathrm{i}\frac{\mu_{0}\omega^{2}}{2p_{z}AB_{\mathrm{in}}}\sum_{j=1}^{M}\frac{\Bigl(\Omega_{j}^{(\mathrm{eig})}\Bigr)^{2}}{\Bigl( J_{j}^{(\mathrm{eig})}+\Delta\Bigr)+\mathrm{i}\Gamma_{j}^{(\mathrm{eig})}},
\end{split}
\end{equation}
with  the $z$-component of the incident x-ray wave vector $p_{z}$, the surface area of the cavity $A$, the cavity field $B_{\mathrm{cav}}$ and incoming field $B_{\mathrm{in}}$, respectively. Furthermore, $J_{j}^{(\mathrm{eig})}$ and $\Gamma_{j}^{(\mathrm{eig})}$ denote the real and imaginary parts of the eigenenergy of the $j$th eigenstate of $\hat{H}_{\rm I}$ calculated for $\Delta=0$ in Eq.~\eqref{Hnuc}. The Rabi frequency of the $j$th eigenstate $\Omega_{j}^{(\mathrm{eig})}$ is obtained from the Rabi frequency of each nuclear layer $\Omega_{l}$  in Eq.~(\ref{Rabi}) by applying a basis transformation. 
The first term on the right-hand side  of Eq.~\eqref{inout} describes the outgoing field created by the electronic scattering of the incoming field, while the latter part accounts for the field resulting from the nuclear dynamics. Taking a closer look, we can see that the contribution of the eigenstates reaches its maximum once the detuning matches the real part of the eigenvalue such that $\Delta + J_{l}^{(\mathrm{eig})}=0$. However, the heights of these maxima strongly depend on the Rabi frequency of each eigenstate in the numerator. Due to the attenuating Pt layers in our system of stacked cavities, the strength of the cavity field $B_{\mathrm{cav}}(z)$ will decrease for deeper cavities and eigenstates concentrated in the uppermost cavities will contribute most to the outgoing field.

The resulting reflectivity curves $\abs{B_{\mathrm{out}}(\Delta)}^2/\abs{B_{\mathrm{in}}}^2$  for the two cavity systems investigated in Fig.~\ref{edge-states}~(b) are shown in  Fig.~\ref{Reflectivity}. Here, we consider that each cavity is illuminated in grazing incidence at its own first resonant angle, corresponding to the first guided mode of the cavity, namely  $\varphi=\SI{2.4157}{mrad}$
for the trivial case and $\varphi=\SI{2.4067}{mrad}$ for the topological one, respectively. 
The reflectivity curve for the topological case resembles a Lorentzian with a constant reflectivity baseline caused by electronic scattering.
Since both edge states have a high localization at the uppermost cavity where the cavity field is the strongest, they have the highest Rabi frequencies of all eigenstates. The direct contributions of each eigenstate to the reflectivity curve are proportional to $\abs{\Omega_{j}^{(\mathrm{eig})}}^{4}$, such that the curve is mostly induced by both edge states while the contributions of the other eigenstates are negligible. In addition, the edge states have nearly equal eigenvalues and the Lorentzians are nearly identical. Thus, the reflectivity of the topological structure resembles the behaviour of a cavity with just a single embedded nuclear layer as seen in Ref. \cite{standardxraynuc}.

\begin{figure}
    \includegraphics[width=0.4\textwidth]{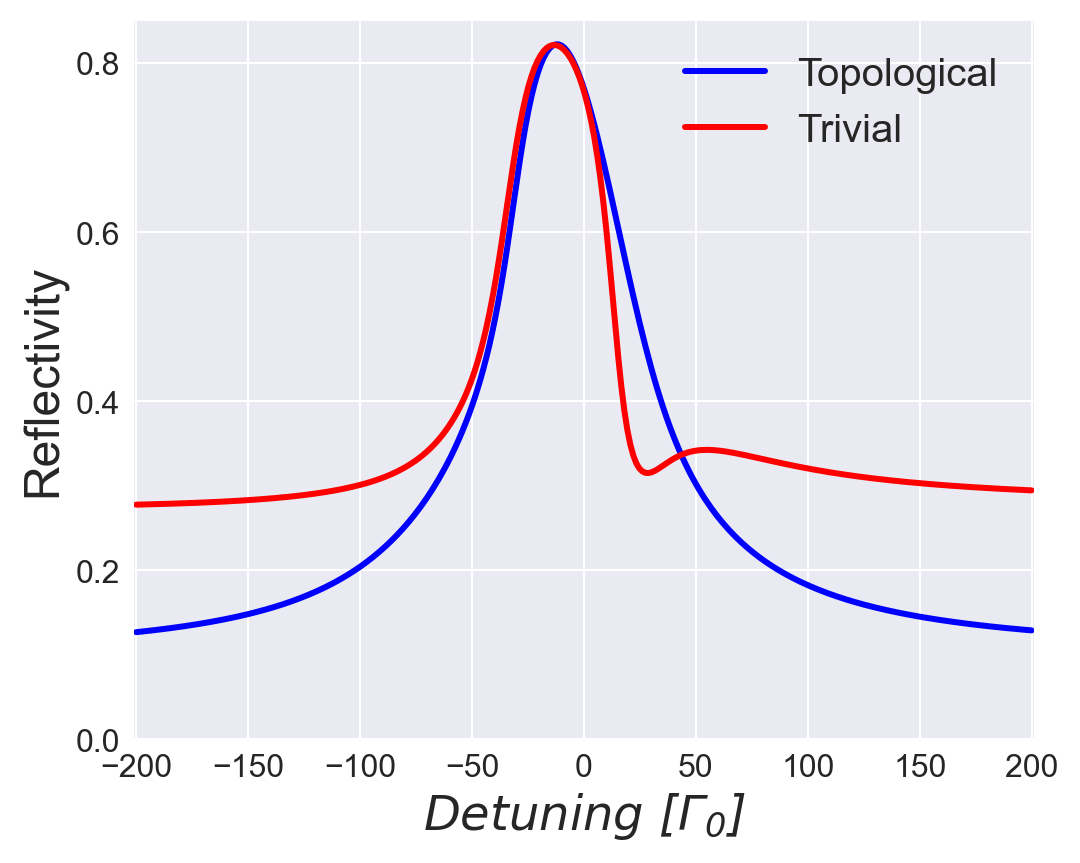}
    \caption{\label{Reflectivity}Reflectivity of a system of ten stacked single-layer cavities in the trivial phase (red line) with a width ratio $d_{\mathrm{v}}/d_{\mathrm{w}}=0.8$ and resonant incidence angle $\varphi=\SI{2.4157}{mrad}$ and topological phase (blue line) with $d_{\mathrm{v}}/d_{\mathrm{w}}=1.4$ and an incidence angle $\varphi=\SI{2.4067}{mrad}$ as a function of detuning.}
\end{figure}

In contrast,  the reflectivity curve of the trivial structure seen in Fig.~\ref{Reflectivity} consists of contributions from many different eigenstates. While the main peak is created by the eigenstates belonging to the upper bundle of eigenvalues shown in Fig.~\ref{edge-states} (a), the significantly smaller peak to the right is due to the eigenstates of the lower bundle of eigenvalues. The dip between the peaks is created by the interference of these eigenstates.
Both the second smaller peak and the local minimum are the main features through which one can differentiate between trivial and topological x-ray cavities. We note that the reflectivity baseline is expected to be different between trivial and topological structures as their different layering leads to  different electronic responses to the incoming x rays.

\emph{Conclusion} ---
Our results show that topological effects can be implemented and probed in thin-film cavities with several layers of $^{57}\mathrm{Fe}$ M\"ossbauer nuclei, where each layer plays the role of a site in a non-Hermitian SSH chain. Multilayer structures with up to $30$ $^{57}\mathrm{Fe}$ layers have been fabricated and probed in x-ray grazing incidence scattering \cite{strongcoupling}. This could be the starting point for further explorations of topology in the x-ray realm, in particular towards x-ray control. An appealing scenario is to combine our findings with the newly demonstrated front coupling regime \cite{FrontcouplingP,FrontcouplingL,Lohse2024}, for instance to implement topological pumping in combination with coupling a second x-ray beam to the front of the cavity.

\emph{Acknowledgments} ---
This research is funded by dtec.bw – Digitalization and Technology Research Center of the Bundeswehr. dtec.bw is funded by the European Union – NextGenerationEU.\\
This work is supported by
the German Science Foundation (Deutsche Forschungsgemeinschaft, DFG) in
the framework of  the Cluster of Excellence on Complexity and Topology
in Quantum Matter ct.qmat (EXC 2147, Project No.
390858490). AP gratefully acknowledges the Heisenberg Program of the
DFG  (Project PA 2508/3-1).

\bibliography{references}

\onecolumngrid
\newpage
\title{Supplementary Material to ``Exploring Topological Effects in Thin-Film X-Ray Cavities''}

\begin{abstract}
In this Supplementary Material, we provide details on computing the cavity field, on the input-output formalism, as well as Hamiltonian matrices in the trivial and topological configurations.
\end{abstract}

\maketitle

\onecolumngrid

\section{Computing the Cavity Field via the dyadic Green's Function}
In the following, we derive the magnetic component of the cavity field $\mathbf{B}_{\mathrm{1D}}(z)$ from Eq. (\ref{Rabi}) in the main text from the dyadic Green's function. For the sake of simplicity, we start by obtaining an expression for the electric field component and subsequently derive the magnetic field one. In general, for linear and isotropic material the electric field generated by a source $\mathbf{F}_{\mathrm{e}}$ reads \cite{Lohse2024}
\begin{equation}\tag{S.1}
    \mathbf{E}(\mathbf{r})=\int \mathbf{G}_{\mathrm{e}}(\mathbf{r},\mathbf{r}')\cdot \mathbf{F}_{\mathrm{e}}(\mathbf{r}')\, d^{3}\mathbf{r}'\,,
\end{equation}
where $\mathbf{G}_{\mathrm{e}}(\mathbf{r},\mathbf{r}')$ denotes the Green's function of the electric field observed at $\mathbf{r}$ with the source location $\mathbf{r}'$. The structures in which we are interested are layered only with respect to the $z$ axis; thus the structure is translationally symmetric in the $(x,y)$ direction. We further assume the source terms to be constant over $y$. So, without loss of generality we can place the plane of incidence of the incoming electromagnetic fields parallel to the $(x,z)$ plane.
As a result, we can neglect the $y$ direction only considering a two-dimensional problem. Considering a purely electric source current
\begin{equation}\tag{S.2}
    \mathbf{F}_{\mathrm{e}}(x,z)=\mathrm{i}\omega\mu_{0}\mathbf{J}_{\mathrm{e}}(x,z)\,,
\end{equation}
with angular frequency $\omega$ and the electric current density $\mathbf{J}_{\mathrm{e}}(x,z)$ yields for the electric field
\begin{equation}\tag{S.3}
\label{electric2}
    \mathbf{E}(x,z)=\mathrm{i}\omega\mu_{0}\int\mathbf{G}(x-x',z,z')\cdot \mathbf{J}_{\mathrm{e}}(x',z')\,dx'\,dz'\, .
\end{equation}
In the case of a plane wave scattering off a layered system in grazing incidence, we assume the source to be situated in the top layer of the structure, i.e., the superstrate. For the source, we consider an electrical current density generating a $\mathrm{p}$-polarized wave coming in under the incidence angle $\varphi$
\begin{equation}\tag{S.4}
    \mathbf{J}_{\mathrm{e}}=\delta(z-z_{\mathrm{src}})e^{\mathrm{i}n_{\mathrm{top}}k\cos(\varphi)}J_{\mathrm{e},0}(\varphi)\mathbf{e}_{\pi}(\varphi)\,,
\end{equation}
with the refractive index of the top layer $n_{\mathrm{top}}$, the position of the source inside the top layer $z_{\mathrm{src}}$ and the $\mathrm{p}$-polarization vector $\mathbf{e}_{\pi}(\varphi)=\cos(\varphi)\mathbf{z}-\sin(\varphi)\mathbf{x}$.
Inserting this current density into the electric field in Eq.~(\ref{electric2}), we obtain the final expression for the scattered field inside the structure
\begin{equation}\tag{S.5}
\label{electrical3}
    \mathbf{E}(x,z)=\frac{2\mathrm{i}n_{\mathrm{top}}\sin(\varphi)E_{0}}{\mu_{\mathrm{top}}}e^{\mathrm{i}n_{\mathrm{top}}k\cos(\varphi)x}\mathbf{G}(p^{\rho}=n_{\mathrm{top}}k\cos(\varphi),z,z_{\mathrm{src}})\mathbf{e}_{\pi}(\varphi)
\end{equation}
where $E_{0}=-\mu_{0}\mu_{\mathrm{top}}J_{\mathrm{e},0}c/(2n_{\mathrm{top}}\sin(\varphi)$) is the amplitude of the incoming electric field, $\mu_{\mathrm{top}}$ is the relative permeability of the top layer and $\mathbf{G}(p^{\rho},z,z_{\mathrm{src}})$ denotes the Green's function, partially Fourier-transformed.
Using Maxwell's equations, we then obtain the magnetic field component from Eq. (\ref{electrical3})
\begin{equation}\tag{S.6}
\label{cavfield}
    \mathbf{B}(x,z)=-\frac{2\mathrm{i}n_{\mathrm{top}}\sin(\varphi)B_{0}}{\mu_{\mathrm{top}}}e^{\mathrm{i}n_{\mathrm{top}}k\cos(\varphi)x}\mathbf{G}(p^{\rho},z,z_{\mathrm{src}})\mathbf{y}
\end{equation}
with a polarization in $\mathrm{y}$ direction and the amplitude of the incoming magnetic field $B_{0}=E_{0}/c$. The magnetic component of the cavity field $\mathbf{B}_{\mathrm{1D}}(z)$ is then obtained through the definition \cite{standardxraynuc}
\begin{equation}\tag{S.7}
    \mathbf{B}(x,z)=\mathbf{B}_{\mathrm{1D}}(z)e^{\mathrm{i}p^{\rho}x}\, .
\end{equation}

\section{Expressing the Input-Output Formalism in the Quasi-Eigenbasis}
We derive an expression for the input-output formalism used in Eq.~(\ref{inout}) for thin-film x-ray cavities in terms of the eigenstates and -energies of the nucleus-nucleus interaction Hamiltonian.
As already established in Eq. (\ref{Hnuc}) the interaction Hamiltonian reads (with $\hbar=1$)
\begin{equation}\tag{S.8}
       \label{Hnucnuc}
    \hat{H}_{ \mathrm{I}}(\Delta) = -\Bigl(\Delta + \mathrm{i}\frac{\Gamma_{0}}{2}\Bigr)\sum_{j=1}^{M}\hat{S}_{j}^{\dagger}\hat{S}_{j}
     -\sum_{j,l=1}^{M}\Bigl(J_{jl}+\mathrm{i}\frac{\Gamma_{jl}}{2}\Bigr)\hat{S}_{j}^{\dagger}\hat{S}_{l}\, ,
\end{equation}
which we express here as a function of the detuning $\Delta$. In the following, we only regard the zero-detuning Hamiltonian $\hat{H}_{ \mathrm{I}}(0)$, as the detuning can then be viewed as a simple eigenenergy shift without any impact on the corresponding eigenstates. Let $\{ |\phi_{j}\rangle\}$ be the set of right eigenstates of the Hamiltonian $\hat{H}_{\mathrm{I}}(0)$ with non-degenerate eigenenergies such that
\begin{equation}\tag{S.9}
\label{lefteig}
    \hat{H}_{\mathrm{I}}(0)|\phi_{j}\rangle=\Bigl( J_{j}^{(\mathrm{eig})}+\mathrm{i}\Gamma_{j}^{(\mathrm{eig})}\Bigr)|\phi_{j}\rangle
\end{equation}
where $J_{j}^{(eig)}$ and $\Gamma_{j}^{(eig)}$ are the real and imaginary parts of the eigenenergy, respectively. Due to the non Hermiticity of the Hamiltonian, $\{|\phi_{j}\rangle\}$ does not form an orthonormal basis as in general the right eigenstates are not orthogonal. However, it is possible to define a biorthogonal set of bases $\{|\phi_{j}\rangle,|\chi_{j}\rangle\}$ in which $|\phi_{j}\rangle$ are linearly independent and therefore form an exact set of basis of the Hilbert space $\mathcal{H}$ \cite{Brody_2014}. Thereby, the states $|\chi_{j}\rangle$ are the left eigenstates of the Hamiltonian.
Although, $\hat{H}_{\mathrm{I}}(0)$ is non-Hermitian, it is complex symmetric meaning $\hat{H}_{\mathrm{I}}(0)=\hat{H}_{\mathrm{I}}^{\mathrm{T}}(0)$ due to the symmetry of the Green's function $G(z_{j},z_{l})=G(z_{l},z_{j})$.
Therefore, our left eigenstates $|\chi_{j}\rangle$ are just the complex conjugated right eigenstates with $\langle\phi_{j}^{\ast}|=(|\phi_{j}\rangle)^{\mathrm{T}}$.
The biorthogonal basis allows us to transform the nuclear-interaction Hamiltonian into its eigenbasis, which yields
\begin{equation}\tag{S.10}
    \hat{H}_{\mathrm{I}}^{(\mathrm{eig})}(0)=\sum_{j=1}^{M}\Bigl(J_{j}^{(\mathrm{eig})}+\mathrm{i}\Gamma_{j}^{(\mathrm{eig})}\Bigr)\hat{S}_{j}^{(\mathrm{eig})\dagger}\hat{S}_{j}^{(\mathrm{eig})}\, ,
\end{equation}
where $\hat{S}_{j}^{(\mathrm{eig})\dagger}=\sum_{l=1}^{M}c_{j}^{(l)}\hat{S}_{l}^{\dagger}$ is the collective nuclear spin-wave operator for the $j$th eigenstate with
\begin{equation}\tag{S.11}
    c_{j}^{(l)}=\frac{\langle\phi_{j}^{\ast}|l\rangle}{\sqrt{\langle\phi_{j}^{\ast}|\phi_{j}\rangle}}\, .
\end{equation}
Whilst in general the eigenstates of the interaction Hamiltonian differ from the effective Hamiltonian $\hat{H}_{\mathrm{eff}} = \hat{H}_{\mathrm D}+\hat{H}_{\mathrm{I}}$ with $\hat{H}_{\mathrm D}$ established in Eq.~(\ref{Hdrive}), we can still utilize them to transform the effective Hamiltonian into what we call the quasi-eigenbasis (QEB). There, at least a sub-space of the Hamiltonian is diagonalized, so that
\begin{equation}\tag{S.12}
    \hat{H}_{\mathrm{eff}}^{\mathrm{QEB}}(\Delta)=\sum_{j=1}^{M}\Bigl(J_{j}^{(\mathrm{eig})}+\Delta+\mathrm{i}\Gamma_{j}^{(\mathrm{eig})}\Bigr)\hat{S}_{j}^{(\mathrm{eig})\dagger}\hat{S}_{j}^{(\mathrm{eig})}+\sum_{j=1}^{M}\Bigl(\Omega_{j}^{(\mathrm{eig})}\hat{S}_{j}^{(\mathrm{eig})\dagger}+\Omega_{j}^{(\mathrm{eig})\ast}\hat{S}_{j}^{(\mathrm{eig})}\Bigr),
\end{equation}
where
\begin{equation}\tag{S.13}
    \Omega_{j}^{(\mathrm{eig})}=\sum_{l=1}^{M}c_{j}^{(l)}\Omega_{l}
\end{equation}
is the Rabi frequency of the $j$th eigenstate. 
The general input-output formalism relates the incoming field into the structure to the outgoing field which we measure at the detector as either  reflectivity or  transmission. For our system, this relation takes the following form in the nuclear-layer basis
\begin{equation}\tag{S.14}
\label{iogeneral}
    \hat{B}_{\mathrm{out}}(z)=\hat{B}_{\mathrm{cav}}(z)+\frac{\mu_{0}\omega_{\mathrm{p}}^{2}}{A}\sum_{j=1}^{M}\sqrt{N_{j}}\mathbf{G}(z,z_{j},\omega_{\mathrm{p}})\cdot\mathbf{m}\hat{S}_{j},
\end{equation}
where the first part describes the classical optical response of the structure and the second part describes the scattering of the field at the nuclear layers.
In the following, we assume that the incoming field is linearly polarized with either s- or p-polarization. For the case of grazing incidence, we can then neglect the vector character of the magnetic field as the polarization gets preserved during the classical scattering at the structure and the scattering at the nuclei. Therefore, both the Green's function $G(z,z',\omega_{\mathrm{p}})$ and the magnetic dipole moment operator $m$ can be treated as scalars, where for the latter its value depends on the chosen polarization while for the former it is approximately independent.\\
Utilizing the formula for the cavity field in Eq. (\ref{cavfield}) and using the fact that $m$ is entirely real valued for $^{57}\mathrm{Fe}$, the Rabi frequency from Eq. (\ref{Rabi}) becomes
\begin{equation}\tag{S.15}
\label{magneticfield}
    \Omega_{j}=-\frac{2\mathrm{i}p_{\mathrm{src}}\sqrt{N_{k}}m}{\hbar}G(z_{j},z_{\mathrm{src}},\omega_{\mathrm{p}})B_{0},
\end{equation}
with the position of the source $z_{\mathrm{src}}$ and the $z$-component of the wave vector in the source layer $p_{\mathrm{src}}$.
 Assuming that we evaluate the outgoing magnetic field at the same position as the source, Eq.(\ref{iogeneral}) can be transformed into
 \begin{equation}\tag{S.16}
    \hat{B}_{\mathrm{out}}=\hat{B}_{\mathrm{cav}}+\mathrm{i}\frac{\mu_{0}\omega_{\mathrm{p}}^{2}\hbar}{2p_{\mathrm{src}}AB_{\mathrm{in}}}\sum_{j=1}^{M}\Omega_{j}\hat{S}_{j}\, .
\end{equation}
Transforming the Rabi frequency and collective spin operators into the quasi-eigenbasis and taking the expectation value $\langle\cdot\rangle$ of the outgoing field with $\langle \hat{S}_{j}^{(\mathrm{eig})}\rangle=S_{j}^{(\mathrm{eig})}$ describing the collective coherence of the $j$th eigenstate yields
\begin{equation}\tag{S.17}
\label{ioqeb}
    B_{\mathrm{out}}(z_{\mathrm{src}})=B_{\mathrm{cav}}(z_{\mathrm{src}})+\mathrm{i}\frac{\mu_{0}\omega_{\mathrm{p}}^{2}\hbar}{2p_{\mathrm{src}}AB_{\mathrm{in}}}\sum_{j=1}^{M}\Omega_{j}^{(\mathrm{eig})}S_{j}^{(\mathrm{eig})}\, .
\end{equation}
We assume that enough time has passed for the collective decoherence to have evolved towards a steady state (meaning $\dot{S}_{k}^{(\mathrm{eig})}=0$). Solving the corresponding Heisenberg equation  leads to the expression
\begin{equation}\tag{S.18}
\label{steadystate}
    S_{j}^{(\mathrm{eig})}=-\frac{\Omega_{j}^{(\mathrm{eig})}}{\Bigl(J_{j}^{(\mathrm{eig})}+\Delta\Bigr)+\mathrm{i}\Gamma_{j}^{(\mathrm{eig})}}\, .
\end{equation}
By inserting Eq.(\ref{steadystate}) back into the input-output formalism in Eq.(\ref{ioqeb}) results in the final expression
\begin{equation}\tag{S.19}
    B_{\mathrm{out}}(z_{\mathrm{src}})=B_{\mathrm{cav}}(z_{\mathrm{src}})-\mathrm{i}\frac{\mu_{0}\omega_{\mathrm{p}}^{2}\hbar}{2p_{\mathrm{src}}AB_{\mathrm{in}}}\sum_{j=1}^{M}\frac{\Bigl(\Omega_{j}^{(\mathrm{eig})}\Bigr)^{2}}{\Bigl(J_{j}^{(\mathrm{eig})}+\Delta\Bigr)+\mathrm{i}\Gamma_{j}^{(\mathrm{eig})}}\, ,
\end{equation}
where $B_{\mathrm{cav}}(z_{\mathrm{src}})=B_{\mathrm{1D}}(z_{\mathrm{src}})-B_{\mathrm{in}}$.

\section{Hamiltonian Matrix for Trivial and Topological Configurations}
Here, we present the Hamiltonian matrices for two different configurations of separating-layer widths for our ten-layer cavity. These configurations correspond to the trivial case with a ratio of $v/w=0.8$ and the topological case with $v/w=1.4$ for which the eigenstates are shown in Fig. \ref{edge-states} (c). 
We denote the nuclear interaction Hamiltonian in the form $\hat{H}_{\mathrm{I}}=\sum_{j,l}(\hat{H}_{\mathrm{I}})_{jl}\,|j\rangle\langle l|$ with $(\hat{H}_{\mathrm{I}})_{jl}=\langle j|\hat{H}_{\mathrm{I}}|l\rangle$, where $|j\rangle$ describes the excited state of the $j$th nuclear layer.
For illustration purposes, we only present the absolute value of each matrix element. 
For the trivial case, this reads
\begin{equation}\tag{S.20}
\left|(\hat{H}_{\mathrm{I}})_{jl}^{\mathrm{triv}}\right|=
\begin{pmatrix}
    22.22 &	21.69 &	9.57 &	9.54 &	4.21 &	4.20 &	1.85 &	1.85 &	0.810 &	0.82\\
    21.69 &	21.79 &	9.40 &	9.38 &	4.14 &	4.13 &	1.82 & 1.82 &	0.80 &	0.81\\
    9.57 &	9.40 & 21.96 &	21.42 &	9.45 &	9.43 &	4.16 &	4.16 &	1.82 & 1.85\\
    9.54 &	9.38 & 21.42 &	21.94 &	9.46 &	9.44 &	4.17 &	4.16 &	1.82 & 1.85\\
    4.21 & 4.14 &	9.45 &	9.46 &	21.94 &	21.40 &	9.44 &	9.43 &	4.13 &	4.20\\
    4.20 & 4.13 &	9.43 &	9.44 &	21.40 &	21.94 & 9.47 & 9.45 & 4.14 &	4.21\\
    1.85 & 1.82 &	4.16 &	4.17 &	9.44 &	9.46 &	21.94 &	21.42 &	9.38 &	9.54\\
    1.85 &	1.82 &	4.16 &	4.16 &	9.43 &	9.45 & 21.42 &	21.96 &	9.40 &	9.57\\
    0.81 &	0.80 &	1.82 &	1.82 &	4.13 &	4.14 &	9.38 &	9.40 & 21.79 &	21.69\\
    0.82 &	0.81 &	1.85 &	1.85 &	4.20 &	4.21 &	9.54 &	9.57 &	21.69 &	22.22
\end{pmatrix}\,,
\end{equation}
and for the topological case
\begin{equation}\tag{S.21}
\left|(\hat{H}_{\mathrm{I}})_{jl}^{\mathrm{topo}}\right|=
\begin{pmatrix}
    40.30 &	11.98 &	11.35 &	3.42 &	3.24 &	0.98 &	0.92 &	0.29 &	0.25 & 0.11\\
    11.98 &	28.56 &	26.60 &	8.00 &	7.58 &	2.29 &	2.16 &	0.67 &	0.60 &	0.25\\
    11.35 &	26.60 &	30.40 &	9.00 &  8.53 &	2.57 &	2.43 &	0.75 &	0.67 &	0.29\\
    3.42 &	8.00 &	9.00 &	29.46 &	27.45 &	8.28 & 7.80 &	2.43 &	2.16 &	0.92\\
    3.24 & 	7.58 &	8.53 &	27.45 &	29.60 &	8.78 & 8.28 &	2.57 &	2.29 & 0.98\\
    0.98 &	2.29 &	2.57 &	8.28 &	8.78 &	29.60 &	27.45 &	8.53 &	7.58 & 3.24\\
    0.92 & 2.16 &	2.43 &	7.80 &	8.28 &	27.45 &	29.46 & 9.00 &	8.00 &	3.42\\
    0.29 &	0.67 &	0.75 &	2.43 &	2.57 &	8.53 &	9.00 &	30.40 &	26.60 &	11.35\\
    0.25 & 0.60 &	0.67 & 2.16 &	2.29 &	7.58 &	8.00 &	26.60 &	28.56 &	11.98\\
    0.11 & 0.25 &	0.29 & 0.92 &	0.98 &	3.24 &	3.42 &	11.35 &	11.98 &	40.30
\end{pmatrix}\,,
\end{equation}
in units of $\hbar\Gamma_{0}$, where $\Gamma_{0}$ is the spontaneous decay rate of a single $^{57}\mathrm{Fe}$ nucleus.
For the calculation of the winding number as presented in Fig. \ref{edge-states} (c) the first and last two columns and rows of the Hamiltonian matrix are neglected. This way, differences in the self coupling across the nuclear layers are minimized such that translational symmetry applies.
\end{document}